\documentclass{aastex631}
\usepackage{array}

\shorttitle{Clustering in GRBs}
\graphicspath{{./}{figures/}}

\begin{document}

\title{Exploring Gamma-Ray Burst Diversity: Clustering analysis of emission characteristics of {\it Fermi} and BATSE detected GRBs}

\correspondingauthor{Shabnam Iyyani}
\email{shabnam@iisertvm.ac.in}

\author[0000-0001-6086-4175]{Nishil Mehta}
\affiliation{Indian Institute of Science Education and Research, Thiruvananthapuram, Kerala, India, 695551}
\affiliation{Laboratoire Lagrange, Observatoire de la Côte d’Azur, Université Côte d’Azur, Nice, France}

\author[0000-0002-2525-3464]{Shabnam Iyyani}
\affiliation{Indian Institute of Science Education and Research, Thiruvananthapuram, Kerala, India, 695551}









\begin{abstract}
Gamma-ray bursts (GRBs) are commonly attributed to the demise of massive stars or the merger of binary compact objects. However, their varied emission characteristics strongly imply the existence of multiple GRB classes based on progenitor types, radiation mechanisms, central engines etc. This study utilizes unsupervised clustering with the Nested Gaussian Mixture Model algorithm to analyze {\it Fermi} and BATSE GRB data, identifying four classes (A, B, C, and D) based on duration, spectral peak, and spectral index, comprising approximately 70\%, 10\%, 3\%, and 17\% of the dataset, respectively. Classes A and B consist of long GRBs, C mainly short GRBs, and class D encompasses both short and long GRBs. Using the spectral index, $\alpha$, for the differentiation of radiation models, it is found that classes B and C align with photospheric emission models, while A and D predominantly show synchrotron radiation characteristics. Short GRBs predominantly exhibit photospheric emission, whereas long GRBs show consistency with synchrotron emission. Overall, 63\% of the total bursts exhibit $\alpha$ profiles indicative of synchrotron emission, with the remaining 37\% associated with photospheric emission. The classes were further examined for their progenitor origins, revealing that classes A and D demonstrate a hybrid nature, while classes B and C are predominantly associated with collapsar and merger origins, respectively. This clustering analysis reveals distinct GRB classes, shedding light on their diversity in radiation, duration and progenitor. 

\end{abstract}



\section{Introduction} \label{sec:intro}
Gamma ray bursts (GRBs) are the brightest transient phenomenon known in the cosmos. The launch of the Burst Transient Spectrometer 
Experiment (BATSE) instrument onboard Compton Gamma ray observatory (CGRO) in 1991 
led to the discovery and spectral study of $2704$ GRBs in the energy range $20$ - $2000$ 
keV, during its 9 years of operation \citep{Fishman2013}. The observations were then further enhanced with the launch of {\it Fermi} gamma ray space telescope in 2008, which observes in a wide energy 
range from 8 keV to 40 MeV by the Gamma ray Burst Monitor (GBM, \citealt{Meegan2009}) and from 30 MeV to 300 GeV using the Large Area Telescope (LAT, \citealt{Atwood2009}). 

The immediate emission from the GRBs, dominantly observed in gamma rays is referred to as the prompt emission. The various observational temporal and spectral characteristics of the prompt emission of GRBs are used for exploring the clustering among GRBs. Broadly, the two types of GRB classes \citep{Kouveliotou1993} refer to two main types of progenitors: merger of binary neutron stars or a neutron star - black hole, which gives rise to largely GRBs of shorter duration ($T_{90}\le 2$s) and harder spectra i.e high spectral peak energies; and core-collapse of massive stars which leads to GRBs of longer duration ($T_{90}\ge 2$s) and softer spectra \citep{MacFadyen1999}. 

The hypothesis of merger of compact objects was confirmed with the concurrent detection of 
GRB170817A along with the gravitational waves \citep{Abbott2017} and the observation of kilonova \citep{170817A_KN_blue,170817A_KN_red}. On the other hand, the hypothesis of the progenitor to be a collapsar \citep{MacFadyen1999} is supported by the concurrent observations of long GRBs with Type Ic/b supernovae \citep{Cano2017} as well as the localisation of the long GRBs to the star-forming regions of the galaxies 
\citep{Perley2016}. 

However, a significant number of GRBs are found to possess temporal and spectral properties which do not fall into these conventional criteria for classes of progenitors based on their $T_{90}$ measurements. 
For example, GRB 200826A \citep{Ahumada_etal_2021,Rossi2022} is a short duration GRB but is associated with stellar core-collapse origin; on the other hand, GRB 211211A \citep{Yang_etal_2022,Troja2022} and GRB 230307A 
\citep{Levan_etal_2023,Dichiara_etal_2023} are long duration GRBs associated with the detection of kilonovae \citep{Dimple_etal_2023}. Furthermore, there have been long duration GRBs such as GRBs 060614 and 060505 
\citep{GRB060614A_2006,Fynbo_etal_2006}  where no supernovae were observed to accompany the GRBs. Thus, $T_{90}$ alone cannot conclusively identify the type of progenitor associated with a GRB and a robust identification of the progenitor is the observation of either a supernova or a kilonova emission. Moreover, the radiation mechanism and the central engine responsible for generating the relativistic jets in GRBs remain elusive. 
Consequently, it is plausible that there may be a greater multitude of classes or subclasses of GRBs originating from diverse progenitors, central engines, radiation processes, and other factors.    

Several studies have been done to find different and definitive classes in GRBs.
Majority of these works are summarised in Table 1 of \cite{Salmon2022}. These attempts were made to identify classes in GRBs based on parameters such as duration, hardness ratio, 
fluences, etc \citep{Rajaniemi2002, Hakkila2003} as well as using the morphology of the GRB light curves \citep{Jespersen2020, Steinhardt2023,Dimple_etal_2023}, 
using various machine learning techniques of unsupervised clustering. 

In this study, we make yet another attempt of unsupervised clustering of GRBs observed by {\it Fermi} and BATSE. Our approach involves utilizing a novel methodology that relies on a minimal set of temporal and spectral characteristics, employing the Nested Gaussian Mixture Modeling technique (NGMM). 
In section \ref{Data}, we present the details of the GRB dataset and the parameters used for the study. The methodology of 
the unsupervised clustering technique adopted in this work is described in section 
\ref{methodology}. In the following section \ref{results}, the clustering results, along with 
the robustness of these classifications, are presented. Finally, in section \ref{discuss}, we 
discuss the implications of these classifications in terms of the radiation processes expected in the prompt emission of GRBs, their duration and progenitor types.

\section{GRB Dataset $\&$ Characteristics}
\label{Data}

The spectral and burst catalogs 
of {\it Fermi}\footnote{https://heasarc.gsfc.nasa.gov/W3Browse/fermi/fermigbrst.html} \citep{Gruber_GBM_spectral_catalog_2014,Fermi_burst_catalog2020,Fermi_GRB_spec_catalog2021} and BATSE\footnote{https://heasarc.gsfc.nasa.gov/W3Browse/all/batsegrb.html}  
\citep{BATSE_catalog_Goldstein_2013} missions are used in 
this study. The GRB spectrum in nature looks non-thermal and is phenomenologically modelled using the 
Band function \citep{Band_etal_1993}. The study of GRB spectra poses significant challenges due to their diverse and unique nature, lacking repetition. Several studies, particularly examining exceptionally 
bright GRBs, have revealed that solely using a Band function is insufficient to fully explain 
the observed spectra or a different spectral model is required, especially when analyzing the time-resolved spectrum with the highest photon count \citep{Ackermann_etal_2013_LATcatalog,Vianello_etal_2018,Iyyani_etal_2015}. However, despite this complexity, when analyzing the time-integrated spectrum 
considering the bursts' temporal evolution and varying fluences, it often aligns predominantly with a Band function. Since the radiation model for GRBs remains an open question and our 
study focuses on conducting a comprehensive analysis of GRB spectral and temporal characteristics for classification purposes, we, therefore, opt to exclusively examine the 
spectral properties of the time-integrated spectrum spanning the entire duration of the burst using the Band function. 

Among the various parameters characterizing the burst and its spectrum, the current classification study uses only three main parameters: 
\begin{itemize}
    \item $T_{90}$: The duration of the burst measured in 50 - 300 keV, during which $90\%$ of the burst fluence is accumulated. 
    \item $E_{peak}$: The peak energy, in the $\nu F_{\nu}$ space, of the best-fit Band function model of the spectrum  of the total burst duration and is measured in keV units.  
    \item $\alpha$: The low energy power law index of the best-fit Band function of the spectrum of the total burst duration.
    Particularly, the GRBs with $\alpha$ values less than $+2$ were only considered in the sample, thereby avoiding outliers. 
\end{itemize}
The spectral parameters of the Band function have been previously employed for GRB classification studies by \cite{Hakkila_etal_2000, Acuner_Ryde2018, Horvath_etal_2019}. However, in contrast to their studies, here we use only the $E_{peak}$ and $\alpha$ spectral parameters in addition to the temporal parameter $T_{90}$.  
The spectral parameters, $E_{peak}$ and $\alpha$ correspond to the {\it Fermi} and BATSE spectra studied in the energy range $\sim 8 \, \rm keV$ - $ 40 \, \rm MeV$ and $20 \, \rm keV - 2 \, MeV$, respectively \citep{Gruber_GBM_spectral_catalog_2014}.
The above three parameters represent the independent and well-constrained parameters of the temporal and spectral behaviour of the GRBs. 
The parameter, the high energy power law index, $\beta$, is avoided as in 
many cases, the high energy emission of the GRB spectrum is not well constrained and many times is consistent with a cutoff \citep{Fermi_GRB_spec_catalog2021}. 

The sample for the study of {\it Fermi} data included GRBs detected between 12 July 2008 and 15 July 2018 
and BATSE data included GRBs observed during the entire operational period between 21 April 1991 and 26 May 2000. The initial sample composed of 2362 {\it Fermi} and 2702 BATSE GRBs. 
The datasets were further filtered such that the GRBs with any one of the parameters missing 
or possesess $\alpha >+2$ (to avoid the extreme outliers) were removed.
These filters resulted in the datasets with a final sample size of 2280 {\it Fermi} and 1959 BATSE GRBs which is used for the study. The classification of GRBs was attempted on the {\it Fermi} and {BATSE} datasets separately. 
The study done by \citet{Fermi_GRB_spec_catalog2021} have shown 
that the distributions of $E_{peak}$ and $\alpha$ of GRBs observed by {\it Fermi} and BATSE are nearly in agreement with each other. This motivated us to further attempt the classification on the combined dataset of {\it Fermi} and BATSE which provides a larger sample size. 

\section{Unsupervised Clustering $\&$ Methodology}
\label{methodology}

In this work, we employ the unsupervised clustering method of Gaussian Mixture Modeling (GMM) in a nested approach of analysing the dataset. A Gaussian mixture model \citep{reynolds2009gaussian} is a probabilistic 
model which assumes that the observed GRBs are a sample of a K number of Gaussian distributions with unidentified 
parameters. Mixture models can be seen as a generalization of k-means clustering 
\citep{Hartigan1979} to include details of the covariance structure of the data as well as the locations of the latent Gaussian centers. 
Thereby, GMM allows to model data with complex patterns and heterogeneity by considering the data to be a composite of multiple Gaussian 
distributions, each with its own mean and variance, thereby allowing to identify multiple modes or clusters in the data. 

The optimal value of K in GMM classification in this study is determined using the Silhouette Coefficient (SC). This metric allows us to 
assess the characteristic, such as how close each point in one cluster is to the other points in the neighboring clusters, thereby determine 
the number of clusters. The Silhouette coefficient is defined as 
\begin{equation}
SC = \frac{b-a}{max(a,b)}
\end{equation}
where $a$ is the mean of the intra-cluster distances and $b$ is the mean of the nearest cluster distances. SC value for different number 
of classes can range between -1 to 1, where values $> +0.5$ and closer to +1 indicate that the classes are well separated and clearly 
distinguishable, while SC = 0 indicates that the classes are not well 
distinguishable, and SC=-1 indicates that means of the clusters are wrongly assigned. 
In this study, the GMM  clustering algorithm is executed using the 
Scikit-learn python package \citep{scikit-learn}. 

\begin{figure}
    \includegraphics[width=\columnwidth, height=10cm]{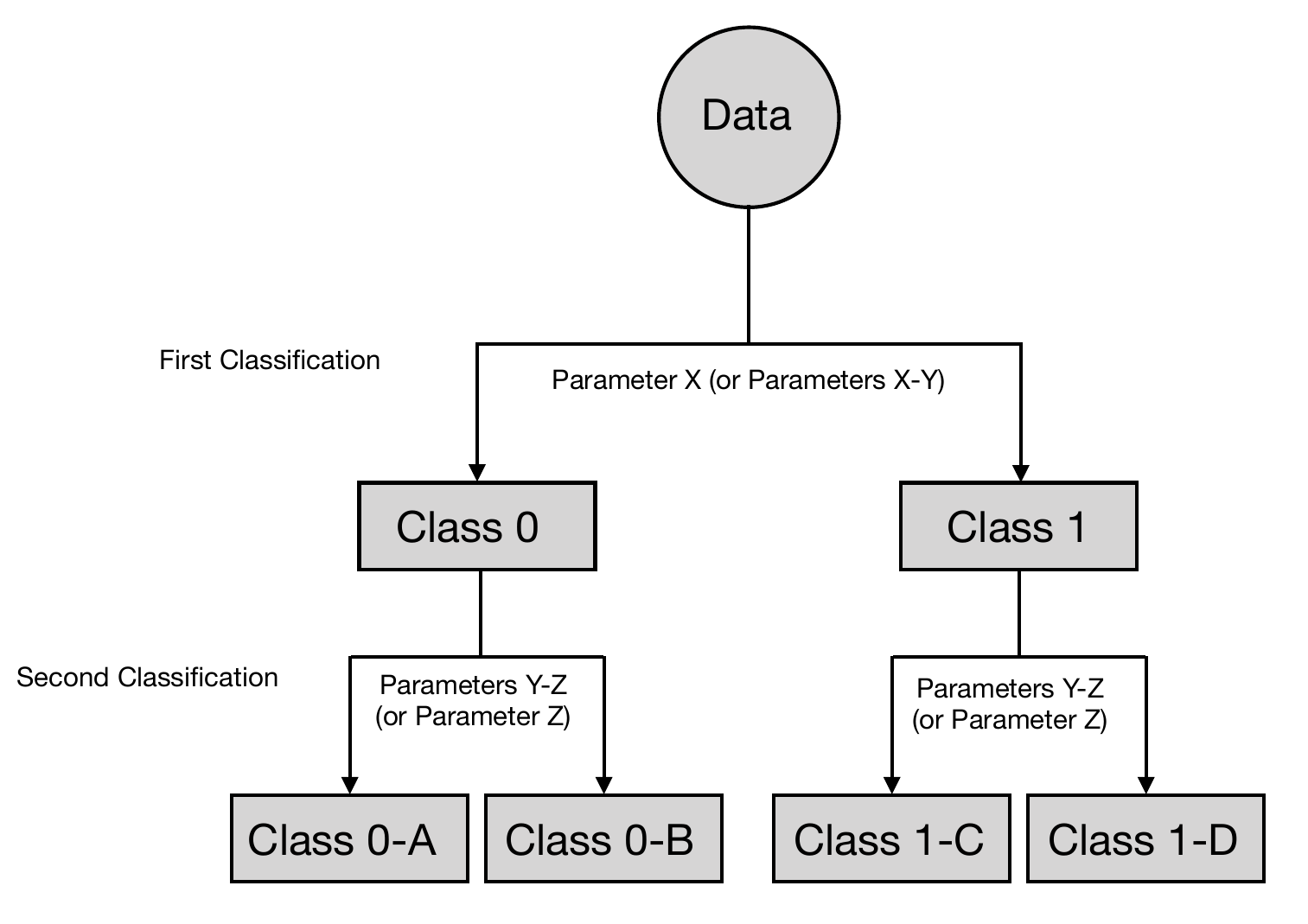}
    \caption{The above flowchart summarises the classification sequence of Nested GMM technique employed on the GRB dataset characterised by three parameters, say X, Y and Z as well as the overall number of classes obtained in this study.}
    \label{flow_chart}
\end{figure}

\subsection{Methodology of clustering}

The technique used for the classification of GRBs is Nested Gaussian mixture modeling (NGMM)\footnote{This technique is named so by the authors.}. In this technique, in the first step, the GRBs are classified using either one or a set of two parameters via the GMM 
method. In this step, the optimal number of classes (K) is determined using the SC. Subsequently, we further classify each of the resulting classes of GRBs in the first step, using either one or a set of two remaining parameters, again, via the GMM method. The number of sub-classes of each class obtained in Step 1 is also determined by the SC. A flow chart outlining the above procedure of the classification is presented in the Fig. \ref{flow_chart}. 

\begin{table}
\resizebox{\columnwidth}{!}{%
\begin{tabular}{ |c|c|c|c| } 
\hline
SNo & Parameter A & Parameter B & Abbreviation of \\
& & & Sequence\\
\hline
1 & $T_{90}$ and $E_{peak}$ & $\alpha$ & TE-A \\ 
2 & $\alpha$ &  $T_{90}$ and $E_{peak}$ & A-TE \\ 
3 & $T_{90}$ & $E_{peak}$ and $\alpha$ &  T-EA\\ 
4 & $E_{peak}$ and $\alpha$ & $T_{90}$& EA-T\\
\hline
\end{tabular}%
}
\caption{The different sequences of GMM classification of GRB dataset using the parameters: $T_{90}$, $E_{peak}$, and $\alpha$.}
\label{table1}
\end{table}

For example, as we attempt a classification of {\it Fermi} detected GRBs using a set of two parameters such as $T_{90}$ and 
$E_{peak}$, we obtain $K=2$ optimal classes with a SC = 0.55, as seen in Fig. \ref{Fermi}I(a). The 2-D histogram of the two classes (F0 and F1) obtained in the first step is shown in Fig. \ref{Fermi}I(b). Each of these classes is then further classified based on the remaining parameter of the dataset, which is $\alpha$. The optimal 
number of sub-classes for both F0 (blue) and F1 (orange) clusters is also found to be 2 each (Fig. \ref{Fermi}Ic) with a SC $> 0.5$. The final result consists of 4 classes of GRBs with their respective histograms of $E_{peak}$, $T_{90}$ and $\alpha$ as shown in Fig. \ref{Fermi}I(d). 

The successfully attempted permutations of this type of sequence of classification of the GRB data, using the three main parameters are listed in Table \ref{table1}.  
We note that attempts of classification of the different datasets using the $E_{peak}$ parameter alone in either the first step or the second step always resulted in ambiguous classifications, which are thereby not listed in the Table \ref{table1}. This indicates that the variability among the GRB classes is not solely influenced by the spectral parameter $E_{peak}$.
This methodology of clustering is applied on {\it Fermi} GRB dataset. The results are discussed in detail in the following section. In addition, the clustering methodology was also applied on $BATSE$ GRB dataset as well as by combining {\it Fermi} and BATSE. The results are presented in Appendix \ref{batse_cluster}
and Appendix \ref{fermi_batse_clustering}, respectively. 

\section{Results}
\label{results}

\begin{figure*}
    \centering
\includegraphics[width=\textwidth]{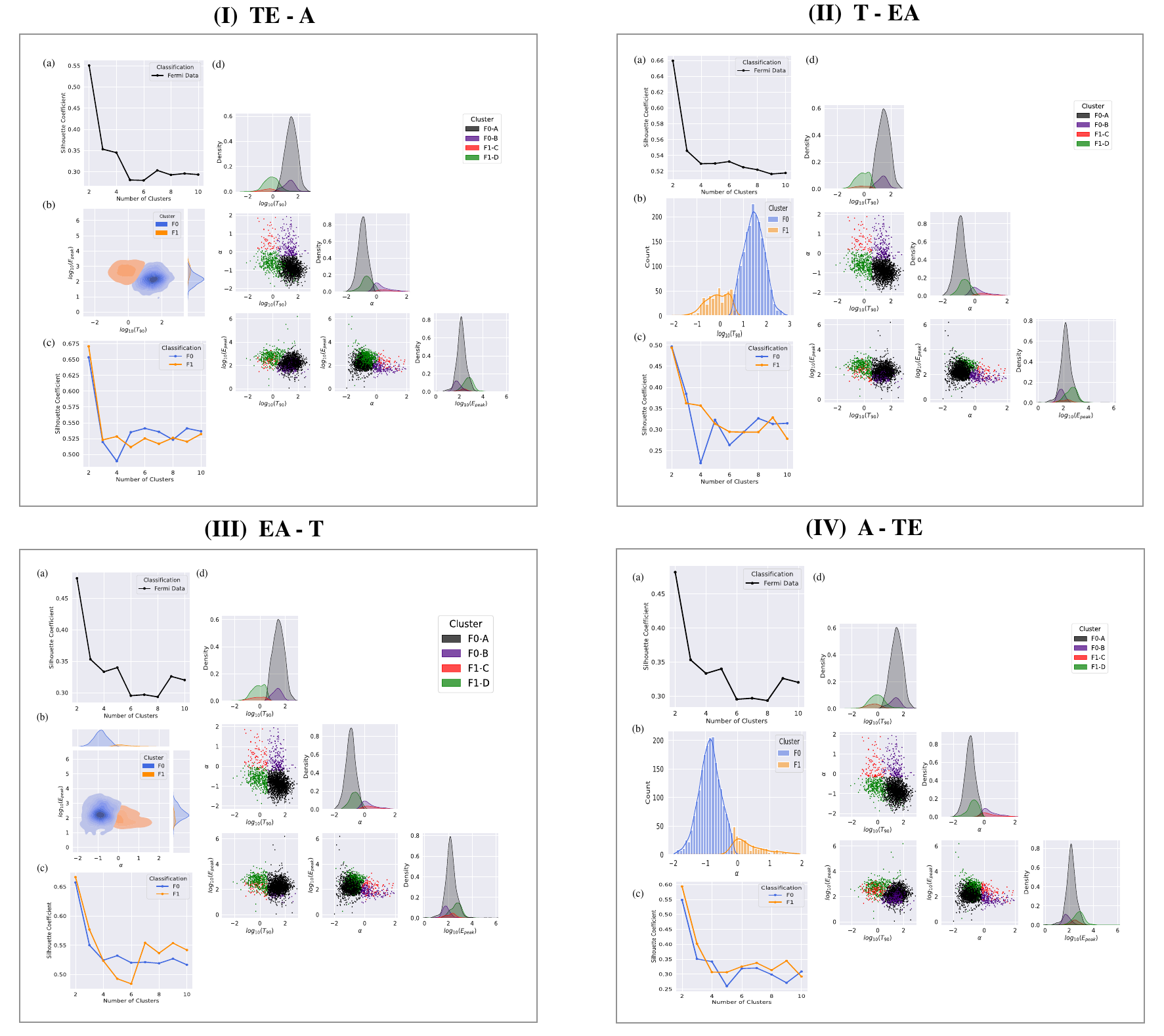}
    \caption{Presented above are the outcomes derived from the NGMM classification applied to the {\it Fermi} GRB dataset. The series of plots labeled as I, II, III, and IV display the results of different classification sequences: TE-A, T-EA, EA-T, and A-TE, respectively. Within each set of plots, (a) and (c) depict the Silhouette Coefficient (SC) obtained during the initial and subsequent phases of the sequence of classifications, respectively. Meanwhile, (b) and (d) portray the KDE illustrating the distribution of properties for the two classes (F0 in blue and F1 in orange) identified in the first step, and all four classes (F0-A in black, F0-B in purple, F1-C in red and F1-D in green) identified in the second step of the classification process, respectively. 
    }
    \label{Fermi}
\end{figure*}

The nested model of GMM clustering (NGMM) is applied on the {\it Fermi} dataset for the various sequences mentioned in the Table \ref{table1}. For the {\it Fermi} dataset, 
the SCs  obtained for classifications both in the first and second steps are found to be $> 0.5$ (Fig. \ref{Fermi}) excepting the first step of classification of the 
sequence EA-T where 
the $SC \approx 0.5$ (subset Fig.\ref{Fermi} III(a) of the sequence EA-T). The higher values of SC $\gtrapprox 0.5$ suggests that the 
classes are clearly distinguishable. In each classification process, 
the SC peaks at K = 2, thereby, resulting in 4 final classes of GRBs. The kernal density estimate (KDE) of the distributions of the parameters characterising the two classes obtained in the first step of classification, labelled as $F0$ and $F1$, and the final four classes, 
labelled as $F0-A$ (black), $F0-B$ (purple), $F1-C$ (red) and $F1-D$ (green), obtained for the sequences TE-A, A-TE, T-EA and EA-T are shown in Fig. \ref{Fermi}. The average and standard deviations of the parameters, 
$E_{peak}$, $T_{90}$ and $\alpha$ of the four classes obtained for the {\it Fermi} dataset in 
the different classification sequences are reported in Table \ref{tab:stats}. In general, 
the classes A, B, C, and D constitute approximately $70\%$, $10\%$, $3\%$, and $17\%$ of the total size of the sample dataset, as indicated in Table \ref{tab:stats}.

Applying the same classification methodology to the BATSE dataset, which comprises 1959 GRBs, yields outcomes as shown in Appendix \ref{batse_cluster} for 
various classification sequences. Within the BATSE dataset, we observe that during the first classification step, an 
optimal number of classes, K=2 (B0 and B1), is obtained for all sequences. The Silhouette Coefficients (SCs) attained for K=2 in the initial 
classification step are generally above 0.5, except in the case of the EA-
T sequence where the SC is 0.4.  In the second classification step, a clear differentiation of K = 2 classes is evident for the EA-T and A-TE 
sequences with SCs above 0.5. However, classification results for both B0 and B1 classes exhibit significant ambiguity in the TE-A sequence, while 
ambiguity is specifically noted for the B1 cluster classification in the T-EA sequence. 

\setlength{\tabcolsep}{8pt} 
\renewcommand{\arraystretch}{2} 

\begin{table}[]

\centering
\resizebox{\columnwidth}{!}{%
\begin{tabular}{|c|c|cccc|} 
\hline 
\textbf{Sequence} & \textbf{Parameters}            & \multicolumn{4}{c|}{\textbf{Fermi [2280]}} \\ 
 \cline{3-6}                                                          
              &                     & A                         & B                         & C                      & D      \\ \hline                
\textbf{TE-A}   & T$_{90}$ (s)          & 29.22$_{-18.72}^{+52.11}$ & 19.71$_{-12.58}^{+34.76}$ & 0.45$_{-0.28}^{+0.76}$ & 0.81$_{-0.59}^{+2.15}$ \\
                                 & E$_{peak}$ (keV)      & 152$_{-92}^{+231}$        & 70$_{-38}^{+83}$          & 210$_{-119}^{+275}$    & 577$_{-397}^{+1273}$ \\  
                                 & $\alpha$              & -0.91 $\pm$ 0.31              & 0.32 $\pm$ 0.49               & 0.69 $\pm$ 0.49            & -0.64 $\pm$ 0.36  \\         
                                 & Fraction              & 0.70                      & 0.10                      & 0.03                   & 0.17        \\ \hline           
\textbf{A-TE}   & T$_{90}$ (s)          & 28.87$_{-18.55}^{+51.87}$ & 20.75$_{-13.23}^{+36.55}$ & 0.5$_{-0.33}^{+1.0}$   & 0.89$_{-0.67}^{+2.61}$ \\
                                 & E$_{peak}$ (keV)      & 150$_{-90}^{+223}$        & 66$_{-34}^{+72}$          & 274$_{-156}^{+359}$    & 622$_{-436}^{+1462}$   \\
                                 & $\alpha$              & -0.9 $\pm$ 0.32               & 0.39 $\pm$ 0.49               & 0.37 $\pm$ 0.53            & -0.72 $\pm$ 0.32        \\   
                                 & Fraction              & 0.71                      & 0.09                      & 0.05                   & 0.15           \\ \hline        
\textbf{T-EA}   & T$_{90}$ (s)          & 29.88$_{-18.76}^{+50.43}$ & 23.34$_{-13.87}^{+34.22}$ & 0.68$_{-0.46}^{+1.46}$ & 0.72$_{-0.49}^{+1.48}$  \\
                                 & E$_{peak}$ (keV)      & 166$_{-108}^{+309}$       & 67$_{-31}^{+56}$          & 144$_{-90}^{+236}$     & 428$_{-278}^{+792}$   \\ 
                                 & $\alpha$              & -0.92 $\pm$ 0.32              & 0.28 $\pm$ 0.52               & 0.64 $\pm$ 0.48            & -0.62 $\pm$ 0.34      \\     
                                 & Fraction              & 0.70                      & 0.10                      & 0.03                   & 0.17    \\ \hline             
\textbf{EA-T}   & T$_{90}$ (s)          & 30.11$_{-18.85}^{+50.37}$ & 23.55$_{-13.66}^{+32.54}$ & 0.67$_{-0.46}^{+1.47}$ & 0.76$_{-0.51}^{+1.57}$ \\
                                 & E$_{peak}$ (keV)      & 165$_{-107}^{+303}$       & 66$_{-34}^{+68}$          & 158$_{-94}^{+234}$     & 432$_{-283}^{+823}$  \\  
                                 & $\alpha$              & -0.92 $\pm$ 0.32              & 0.33 $\pm$ 0.5                & 0.49 $\pm$ 0.51            & -0.65 $\pm$ 0.33      \\     
                                 & Fraction              & 0.70                      & 0.09                      & 0.04                   & 0.17         \\ \hline         
\end{tabular}%
}
\caption{Tabulated are the mean and standard deviation of $T_{90}$, $E_{peak}$, and $\alpha$ parameters for the four clusters, along with their respective sizes as fractions of the dataset. These results are obtained from the various classification sequences applied to the {\it Fermi} dataset. 
The sample size of GRBs considered in the dataset is denoted by the number in the square brackets in the initial row of the table.}
\label{tab:stats}
\end{table}

The identical classification methodology is further extended to the merged dataset of {\it Fermi} and BATSE, encompassing 4239 GRBs. The 
outcomes are detailed in Appendix \ref{fermi_batse_clustering}. A clear differentiation into two classes, 
denoted as C0 and C1, emerges during the initial classification step, with Silhouette Coefficients (SCs) exceeding $0.5$, barring the EA-T sequence where the SC is approximately $0.45$. In the ensuing 
classification step across all sequences (C0 and C1), each yields a division into two classes, with SCs surpassing 0.5, except for C1 within the T-EA sequence, where the classification enters a state of ambiguity. 
The average and standard deviations of the parameters, $E_{peak}$, $T_{90}$ and $\alpha$ of the four classes obtained for the BATSE and merged datasets for the successful different classification sequences are reported in Table \ref{tab:stats_appendix} in the appendix.   

The classes A and B exclusively encompass the category of longer-duration GRBs ($T_{90} > 2$ s), characterized by $T_{90}$ values in the range of several tens of seconds. While class A features soft 
$E_{peak}$ values of a few hundred keV and exhibits softer $\alpha$ parameter around $-0.9$. In contrast, class B comprises of GRBs with relatively even softer $E_{peak}$ values, typically several tens of keV, 
accompanied by a steeper $\alpha$ value greater than $+0.2$.

While class C almost composes of short GRBs i.e., $T_{90} < 2 \, s$, class D predominantly (three quarters) represents GRBs of shorter durations. Within this category, class C is distinguished by its soft $E_{peak}$ values, typically in the few hundred keV range, 
and its $\alpha$ parameter is harder by being greater than $+0.3$. In contrast, class D is representative of both long and short GRBs, possessing higher $E_{peak}$ values hovering around 500 keV, and exhibits softer $\alpha$ below $-0.6$.   

Interestingly, we note that the final four clusters obtained in all the different sequences of classifications on all the three different GRB samples have similar characteristics (Table \ref{tab:stats} and Table \ref{tab:stats_appendix}). Class A and B 
GRBs, typically long GRBs with soft $E_{peak}$ values, differ in that Class A has a soft $\alpha$, whereas Class B has a hard $\alpha$. On 
the other hand, class C and D GRBs, broadly representing short GRBs, have differences too. Class C has softer $E_{peaks}$ similar to typical 
long GRBs (class A) but with very hard $\alpha$, while class D has hard $E_{peak}$ values and relatively softer $\alpha$.

\section{Discussions - Interpretation of Clusters}
\label{discuss}
In the current study, categories of GRBs are determined 
based on their spectral properties and $T_{90}$.
Here, we further investigate these clusters from different perspectives: (i) analyzing the radiation process primarily using the $\alpha$ parameter as a diagnostic along with the duration of the bursts and (ii) 
considering known observations of kilonova or supernova associated with GRBs to shed light on the progenitors of the GRBs of these classes.

\subsection{Types of Radiation Processes}
The fundamental radiation process responsible for the prompt gamma-ray emission in GRBs remains largely an
unresolved mystery. Within the conventional framework of the GRB fireball model, there are two primary competing radiation models: 
photospheric emission \citep{Ryde2004,Ryde&Pe'er2009,Ryde2011,Iyyani2013,Lundman2013,Iyyani_etal_2015} and synchrotron radiation \citep{Rees_Meszaros1994,Tavani1996,Papathanassiou1996,Beniamini_etal_2018}. In the case of a non-
dissipative photosphere, the emission is expected to resemble a blackbody (with $\alpha = +1$), although when considering 
geometric effects, $\alpha$ can become softer than +1 or equal to +0.5 \citep{Pe'er2008,Beloborodov2011,Lundman2013}. When modeled using an empirical function like a Band function, its limitation 
to capture the true curvature of the non-dissipative photospheric 
spectrum, is found to further soften the estimated $\alpha$ such that $-0.4 \leq \alpha \leq 0.0$ \citep{Acuner_etal_2020}. Additionally, accounting for the temporal evolution of the blackbody emission within the time-integrated spectrum as well as the dissipation of kinetic/ Poynting flux below the photosphere can further soften the observed $\alpha$ of the spectrum emitted from the photosphere \citep{Giannios2008,Ahlgren_etal_2015}.

Regarding synchrotron emission, if the electrons undergo cooling within the dynamical time (the duration it takes for the energized electrons to cross the shock region), the resulting spectrum is termed "fast 
cooling synchrotron" \citep{Sari1998}. When considering the temporal evolution of the spectrum, this can lead to observed time-integrated spectra having an 
$\alpha$ value of less than or equal to -1.5. Conversely, if the electrons take longer than the dynamical time to cool, the spectrum is 
called "slow cooling synchrotron" \citep{Sari1998}, and with temporal evolution considered, the observed time-integrated spectra may have an $\alpha$ 
value of less than or equal to -0.67 (also, refer \citealt{Burgess_etal_2015}). Typically, it is considered that spectra with $\alpha$ values greater than -0.67 cannot be produced by synchrotron radiation and is referred to as 'line of death' of synchrotron radiation \citep{Preece_etal_1998}. As such, 
$\alpha$ values greater than -0.67 are usually attributed to cases potentially originating from the photosphere emission models. However, it is worth noting that alternative models of modifications to the conventional synchrotron radiation are present \citep{Granot_etal_2000,Burgess_etal_2020}.

As the spectral parameter $\alpha$ plays a pivotal role in distinguishing between potential radiation models, we, therefore, employ a 
straightforward approach to assess each class of GRBs identified in our classification process. Our goal is to determine the 
consistency of these classes with different emission categories: photosphere emission (depicted in purple), fast cooling synchrotron (shown in red), or slow cooling synchrotron 
(represented in green) using the criteria $\alpha > -0.67$, $\alpha \leq -1.5$, and $-1.5 < \alpha \leq -0.67$, respectively. In Fig. \ref{fig:sum}, we present the results for the {\it Fermi} GRB dataset and the TE-A classification sequence. 
Additionally, the left side of the grouped bar chart illustrates the proportion of GRBs classified as either long (in blue) or short (in orange) based on their $T_{90}$ durations greater than or less than 2 seconds respectively. We, however, point out that in the current interpretation of the classes, $T_{90}$ is not considered as the indicator of the progenitor of the GRB in the conventional sense. 

In their study, \citet{Acuner_Ryde2018} reported the outcomes of clustering analysis for gamma-ray bursts (GRBs) detected by the {\it Fermi} satellite. They employed a Gaussian Mixture Modeling 
approach that incorporated five parameters, encompassing the high-energy spectral index ($\beta$) and energy fluence, alongside the 
parameters examined in our research and found 5 classes of GRBs which were further interpreted using characteristics such as minimum variability timescale, smoothness parameter, spectral width, redshift etc. Their findings indicated that approximately two-thirds of total bursts were attributed to photospheric emission, and the predominantly short GRB class exhibited consistency with photospheric emission properties.

Upon examination of clusters obtained in this study, we find that Class A consists entirely of long GRBs, with approximately $75\%$ of them aligning with the characteristics of slow cooling synchrotron emission. Only about $5\%$ match the profile 
of fast cooling synchrotron emission, while roughly $20\%$ are consistent with photospheric emission. In contrast, 
Class B comprises solely long GRBs, all of which are in line with  photospheric emission. Class C is dominantly composed of short GRBs, 
all indicating photospheric emission. Meanwhile, class D displays a hybrid nature in terms of both duration and radiation process. Class D contains approximately $25\%$ long GRBs, with the majority exhibiting characteristics of slow cooling synchrotron 
radiation, followed by photospheric emission, and a very small fraction corresponding to fast cooling synchrotron radiation. For short GRBs in Class D, majority are consistent with photospheric emission, followed by slow cooling, with the least matching being the fast cooling synchrotron radiation. 
 
Upon analyzing the clusters, we observe that a significant portion of long GRBs, approximately $69\%$, predominantly exhibits characteristics in line with synchrotron emission. 
This percentage encompasses both fast and slow cooling synchrotron spectra. 
Conversely, short GRBs predominantly align with photospheric emission, constituting approximately $66\%$ of the observed cases. This observation implies that, in general, the prompt emission of short GRBs 
primarily originates from the photosphere, corroborating the findings presented by \citet{Acuner_Ryde2018,Dereli_etal_2020,Iyyani_Sharma2021}. 

Furthermore, it's worth highlighting that among the identified clusters, the prominent Class A constitutes a substantial portion of the {\it Fermi} dataset, accounting for 70\%. While the class comprises 
exclusively of long-duration GRBs, its hybrid nature arises from the presence of GRBs with diverse origins of prompt radiation wherein 
approximately 79\% of the GRBs exhibit consistency with synchrotron emission. Conversely, Class D displays a hybrid nature encompassing both duration and radiation process characteristics. 
Our current analysis further reveals 
that 63\% of the total bursts, exhibits $\alpha$ profiles consistent with synchrotron emission, while only 37\% are attributed to photospheric emission. Nevertheless, it is 
essential to acknowledge that various advanced models exist within the realms of both 
synchrotron \citep{Ghisellini1999,Granot_etal_2000,Asano_etal_2009,Daigne2011,Zhang&Yan2011,Uhm_Zhang2014} and photospheric emissions \citep{Peer&Waxman2004,Beloborodov2011,Giannios2012, Beloborodov2013,Lundman2013}, capable of generating spectra that deviate from the $\alpha$ criteria used for assessing radiation processes as mentioned above. 

\subsection{Types of Progenitors}
The two primary progenitor types of GRBs, which are observationally well-established, are collapsars and binary neutron star mergers. By utilizing the prompt 
emission properties ($\alpha$, $E_{peak}$ and $T_{90}$) of a known set of {\it Fermi} detected GRBs (Table \ref{GRBlist}) with associated supernova \citep{Kovacevic_etal_2014} or 
kilonova emissions \citep{170817A_KN_blue,Kilonova_search_2020} until December 2023, we determine the class to which the GRB most likely to belong\footnote{We have included the tentative associations of supernova and kilonova emissions reported with respect to {\it Fermi} detected GRBs in this study.}. Assuming the 
$\alpha$, $E_{peak}$, and $T_{90}$ parameters of the prompt emission of GRB are independent, we ascertain the probability density of each parameter value within the determined Gaussian 
model of the class. Then, we calculate the combined likelihood of the GRB being part of that class by multiplying these probability densities. By comparing the 
likelihood of the GRB belonging to each class, we assign the GRB to the class with the highest likelihood.

\begin{table}
\resizebox{\columnwidth}{!}{%
\begin{tabular}{ |c|c|c|c|c|c|c| } 
\hline
SNo & Supernova - GRB &  Reference & Class & Kilonova - GRB &  Reference & Class \\
\hline
1 & GRB 090320B & \cite{Kovacevic_etal_2014} & Class A & GRB 080905A & \cite{Kilonova_search_2020} & Class C \\ 
2 & GRB 090426B & \cite{Kovacevic_etal_2014} & Class B & GRB 100206A & \cite{Kilonova_search_2020} & Class D \\ 
3 & GRB 090618  & \cite{Kovacevic_etal_2014} & Class A & GRB 150101B & \cite{Kilonova_search_2020} & Class C \\ 
4 & GRB 091127A & \cite{Kovacevic_etal_2014} & Class A & GRB 150424A & \cite{Kilonova_search_2020} & Class A \\
5 & GRB 101219B & \cite{Kovacevic_etal_2014} & Class B & GRB 160821B & \cite{Kilonova_search_2020} & Class D \\
 & & & & & \cite{160821B_kilonova} & \\
6 & GRB 110911A & \cite{Kovacevic_etal_2014} & Class B & GRB 170817A & \cite{170817A_KN_blue} & Class A \\
& & & & & \cite{170817A_KN_red} & \\
7 & GRB 111228A & \cite{Kovacevic_etal_2014} & Class A & GRB 211211A & \cite{211211A_KN} & Class A \\
8 & GRB 130215A & \cite{Kovacevic_etal_2014} & Class B & GRB 230307A & \cite{230307A_KN} & Class A \\
9 & GRB 130427A & \cite{Kovacevic_etal_2014} & Class A & & & \\
10 & GRB 130702A & \cite{Kovacevic_etal_2014} & Class A & & & \\
11 & GRB 200826A & \cite{Ahumada2021} & Class D & & &  \\
12 & GRB 201015A & \cite{Belkin_etal_2024} & Class B & & &  \\
\hline
\end{tabular}%
}
\caption{The list of {\it Fermi} detected GRBs with known supernova or kilonova observations and their associated GRB classes.}
\label{GRBlist}
\end{table}

\begin{figure}
\centering
\includegraphics[width=\columnwidth]{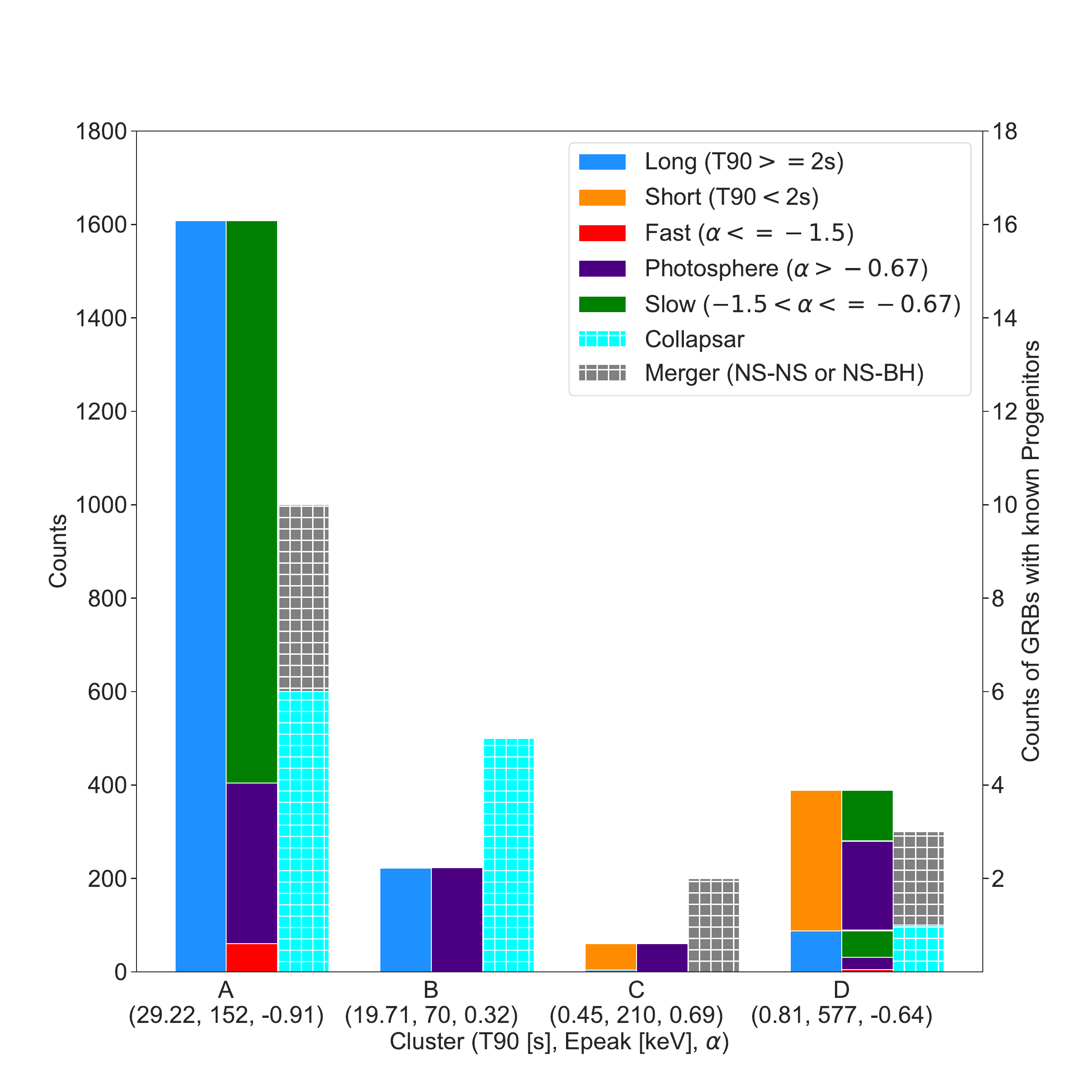}
\caption{The grouped bar chart depicted above illustrates the TE-A NGMM classification results when applied to the {\it Fermi} dataset. In each class, the left bar displays 
the distribution of GRB durations, with blue indicating long-duration GRBs and orange representing short-duration 
ones. Meanwhile, the middle bar showcases the distribution of potential classic emission mechanisms among the GRBs in 
each class. Among these, the red and green bars symbolize the fast and slow cooling synchrotron emission models, 
while the purple bar represents the photosphere emission model. In each class, the right bar (hatched) displays the progenitor types with cyan representing collapsar while grey representing merger of binary neutron star - neutron star or neutron star - black hole. The average values of $T_{90}$, $E_{peak}$ and $\alpha$ are mentioned below of each cluster's label. 
}
\label{fig:sum}
\end{figure}

The bar plot presented in Figure \ref{fig:sum} summarizes the outcomes of associating 20 {\it Fermi} GRBs with known progenitor associations, based on detections of either 
supernovae (cyan hatched) or kilonovae (grey hatched). Classes A and D demonstrate a mixed nature, encompassing 
GRBs originating from both collapsars and the merger of binary neutron stars or neutron star-black hole systems. Specifically, Class A comprises 6 collapsar and 4 merger events, while Class D includes 1 collapsar and 2 merger 
events. Conversely, classes B and C consist solely of 5 collapsar and 2 kilonova events, respectively. Based on the limited data, this analysis indicates that classes A 
and D display hybrid characteristics regarding progenitors, whereas classes B and C tends to be 
exclusively associated with collapsar and merger origins, respectively.

\section{Summary}

In summary, the study examines three parameters ($T_{90}, \, E_{peak}, \, \alpha$) characterising the temporal and spectral properties of the GRBs to identify clustering among GRB datsets of {\it Fermi} and BATSE observations. We introduce for the first time the unsupervised Nested Gaussian Mixture Modelling (NGMM) classification technique which 
have identified four potential distinct classes of GRBs among both {\it Fermi} and BATSE GRB datasets. This 
sequential clustering technique is distinctly different from the conventional Gaussian Mixture Modelling classification technique where the clustering is attempted on the entire chosen parameter space at once. 
Notably, all sequences in NGMM does not result in successful clustering. Attempts to classify the GRB datasets using the $E_{peak}$ parameter alone in either the first or the second step consistently led to 
ambiguous classifications. This reveals that the spectral parameter $E_{peak}$ alone does not drive the variability among the GRB classes. Furthermore, the same methodology of clustering has been applied on 
different datasets such as {\it Fermi}, BATSE and {\it Fermi} + BATSE resulting in successful clustering with four classes exhibiting similar characteristics which affirms the robustness of the methodology of clustering. 

Class A and B type GRBs are typically long duration GRBs with soft $E_{peak}$ values, differing in that Class A has a soft $\alpha$, whereas Class B has a harder $\alpha$. On the other hand, class C type GRBs are dominantly of shorter duration, while D type GRBs are a 
mixture of long and short duration. Class C has softer $E_{peaks}$ similar to typical long GRBs (class A) but with very hard $\alpha$, while class D has brighter $E_{peak}$ values and 
relatively softer $\alpha$. Using the $\alpha$ parameter for the differentiation of radiation models, it is evident that classes B and C align 
with photospheric emission models, while classes A and D predominantly exhibit spectral profiles consistent with slow cooling synchrotron radiation, with 
some contributions from photospheric and fast cooling synchrotron emissions. Furthermore, we find that short GRBs are dominantly photospheric emission 
dominated while long GRBs are relatively more consistent with synchrotron emission. Overall, nearly $63\%$ of the bursts exhibited spectral profiles consistent with synchrotron emission. 
The obtained classes were further investigated in terms of their progenitor orgins by identifying the association of the {\it Fermi} detected GRBs with known progenitors. With the 
available limited data, this analysis suggests that classes A and D exhibit hybrid nature in terms of progenitors, while classes B and C tend to be associated exclusively to collapsar and merger origins, respectively. 
Thus, this study provides insights into the 
dominant radiation processes, duration and progenitors among these classes and sheds light on the complex and fascinating nature of the diversity among the GRBs. 

\newpage
\begin{acknowledgments}
S.I. is supported by the DST INSPIRE
Faculty Scheme (IFA19-PH245) and
SERB SRG grant (SRG/2022/000211). This research has
made use of data obtained from the High Energy Astrophysics
Science Archive Research Center (HEASARC), provided by
NASA’s Goddard Space Flight Center. 
\end{acknowledgments}

\appendix
\section{Clustering in BATSE GRB dataset}
\label{batse_cluster}
\begin{figure*}
    \centering
\includegraphics[width=\textwidth]{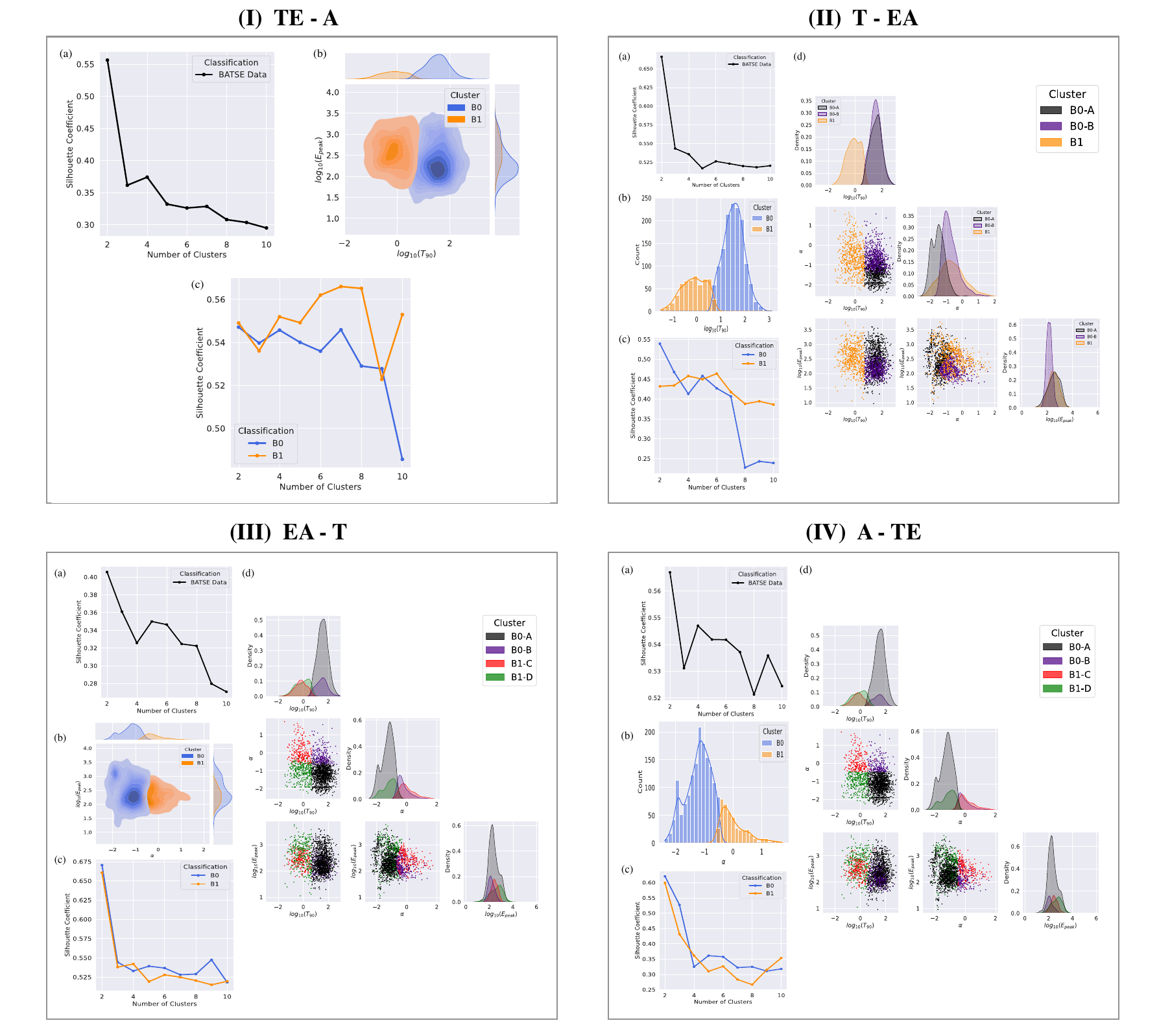}
    \caption{Displayed above are the outcomes of the NGMM classification applied to the BATSE GRB dataset. The set of plots labeled I, II, III, and IV showcase the results of distinct classification sequences: TE-A, T-EA, EA-T, and A-TE, correspondingly. Within each plot set, (a) and (c) exhibit the Silhouette Coefficient (SC) achieved during the initial and subsequent steps of the sequence, respectively. While (b) and (d) illustrate the Kernel Density Estimation (KDE) representing the distribution of properties for the two classes identified in the first step, and for all four classes (B0-A in black, B0-B in purple, B1-C in red and B1-D in green) identified in the second step of the classification process, respectively. However, in sequence TE-A, there is no (d) plot as the second step of classification resulted in ambiguous classes. Similarly, in sequence T-EA, since the second step of classification of B1 resulted in ambiguous classes, (d) plot showcases only the sub-classes of B0: B0-A (break) and B0-B (purple), and B1 (orange) class.}
    \label{batse}
\end{figure*}

The results obtained for the NGMM clustering technique when applied on the BATSE GRB dataset is presented in Figure 
\ref{batse}. The means, the standard deviations and the respective sizes in terms of the fraction of the dataset, for the four classes are listed in Table \ref{tab:stats_appendix}. 

The first step of classification in all the sequences resulted in two classes: B0 (blue) and B1 (orange). The further classification of these clusters were fully successful only in two sequences: EA-
T and A-TE. We note that the classification in the second step of the sequence TE-A resulted in ambiguous clusters as evident in 
Figure \ref{batse}I (c).  Therefore, the corresponding row is left vacant in the Table \ref{tab:stats_appendix}. We also note that in the classification attempt of the class B1 in the sequence T-EA also resulted in ambiguous number of clusters and therefore Figure \ref{batse}II (d), shows the two sub-classes obtained from successful classification of B0, which are the class A in black and class B in purple, along with class B1 (orange). 

\setlength{\tabcolsep}{12pt} 
\renewcommand{\arraystretch}{2} 

\begin{table*}[]

\centering
\resizebox{\textwidth}{!}{%
\begin{tabular}{|c|c|cccc|cccc|}
\hline
\textbf{Sequence} & \textbf{Parameters} & \multicolumn{4}{c|}{\textbf{BATSE [1959]}}                                                              & \multicolumn{4}{c|}{\textbf{Combined (Fermi+BATSE) [4239]}}                                              \\
                                  & \multicolumn{1}{l|}{} 
                                 & A                         & B                         & C                      & D                      & A                         & B                         & C                      & D                      \\ \hline
\textbf{TE-A}   & T$_{90}$ (s)          
& & & &
& 30.69$_{-19.65}^{+54.63}$ & 21.36$_{-13.82}^{+39.11}$ & 0.49$_{-0.31}^{+0.83}$ & 0.74$_{-0.51}^{+1.65}$ \\
                                 & E$_{peak}$ (keV)      
                                 & & & &
                                 & 179$_{-110}^{+285}$       & 85$_{-47}^{+104}$         & 228$_{-109}^{+209}$    & 499$_{-325}^{+929}$    \\
                                 & $\alpha$              
                                 & & & &
                                 & -1.02 $\pm$ 0.42              & 0.32 $\pm$ 0.53               & 0.57 $\pm$ 0.46            & -0.77 $\pm$ 0.48           \\
                                 & Fraction              
                                 & & &  &
                                 & 0.70                      & 0.07                      & 0.04                   & 0.19                   \\ \hline
\textbf{A-TE}   & T$_{90}$ (s)          
& 34.36$_{-21.74}^{+59.21}$ & 26.17$_{-16.35}^{+43.55}$ & 0.62$_{-0.4}^{+1.12}$  & 0.78$_{-0.55}^{+1.87}$ & 30.78$_{-19.71}^{+54.84}$ & 20.13$_{-13.23}^{+38.61}$ & 0.5$_{-0.31}^{+0.84}$  & 0.78$_{-0.55}^{+1.88}$ \\
                                 & E$_{peak}$ (keV)      
                                 & 225$_{-136}^{+345}$       & 132$_{-61}^{+113}$        & 294$_{-142}^{+273}$    & 466$_{-301}^{+847}$    & 177$_{-108}^{+277}$       & 83$_{-45}^{+98}$          & 257$_{-124}^{+239}$    & 530$_{-350}^{+1027}$   \\
                                 & $\alpha$              
                                 & -1.23 $\pm$ 0.43              & -0.04 $\pm$ 0.4               & 0.13 $\pm$ 0.49            & -1.11 $\pm$ 0.44           & -1.02 $\pm$ 0.42              & 0.36 $\pm$ 0.45               & 0.4 $\pm$ 0.49             & -0.83 $\pm$ 0.45           \\
                                 & Fraction              
                                 & 0.62                      & 0.10                      & 0.11                   & 0.17                   & 0.70                      & 0.07                      & 0.05                   & 0.18                   \\ \hline
\textbf{T-EA}   & T$_{90}$ (s)          
& 35.38$_{-22.04}^{+58.47}$ & 33.75$_{-20.76}^{+53.96}$ 
& & 
& 32.31$_{-20.29}^{+54.53}$ & 25.12$_{-14.85}^{+36.33}$ 
& &
\\
                                 & E$_{peak}$ (keV)      
                                 & 359$_{-236}^{+689}$       & 147$_{-58}^{+96}$         
                                 & & 
                                 & 197$_{-127}^{+356}$       & 92$_{-40}^{+71}$          
                                 & &
                                 \\
                                 & $\alpha$              
                                 & -1.49 $\pm$ 0.4               & -0.74 $\pm$ 0.49               
                                 & &     
                                 & -1.07 $\pm$ 0.4               & 0.03 $\pm$ 0.55               
                                 & &    
                                 \\
                                 & Fraction              
                                 & 0.31                      & 0.39                     
                                 & &           
                                 & 0.65                      & 0.11                      
                                 & &
                                 \\ \hline
\textbf{EA-T}   & T$_{90}$ (s)          
& 35.4$_{-21.88}^{+57.27}$  & 30.16$_{-18.85}^{+50.24}$ & 0.66$_{-0.42}^{+1.2}$  & 0.83$_{-0.58}^{+1.99}$ & 32.25$_{-20.1}^{+53.35}$  & 23.54$_{-14.1}^{+35.17}$  & 0.62$_{-0.4}^{+1.15}$  & 0.78$_{-0.54}^{+1.71}$ \\
                                 & E$_{peak}$ (keV)      
                                 & 239$_{-148}^{+385}$       & 150$_{-68}^{+125}$        & 269$_{-124}^{+231}$    & 407$_{-279}^{+886}$    & 189$_{-120}^{+331}$       & 88$_{-44}^{+90}$          & 208$_{-109}^{+227}$    & 406$_{-268}^{+786}$    \\
                                 & $\alpha$              
                                 & -1.28 $\pm$ 0.41              & -0.21 $\pm$ 0.41              & 0.06 $\pm$ 0.51            & -1.14 $\pm$ 0.41           & -1.03 $\pm$ 0.42              & 0.31 $\pm$ 0.47               & 0.4 $\pm$ 0.48             & -0.82 $\pm$ 0.44           \\
                                 & Fraction              
                                 & 0.57                      & 0.14                      & 0.13                   & 0.16                   & 0.69                      & 0.07                      & 0.06                   & 0.18                   \\ \hline
\end{tabular}%
}
\caption{Tabulated are the mean and standard deviation of $T_{90}$, $E_{peak}$, and $\alpha$ parameters for the four clusters, along with their respective sizes as fractions of the dataset. These results were obtained from various classification sequences applied to the BATSE, and {\it Fermi} + BATSE datasets. The sample size of GRBs considered in each dataset is denoted by the number in square brackets in the initial table row. Note: The vacant cells in the table indicate specific scenarios. In the case of the TE-A classification sequence applied to the BATSE dataset, the second step did not provide robust constraints for the classes. Likewise, when employing the T-EA classification sequence on both the BATSE and Fermi + BATSE datasets, it was not possible to adequately define subclasses for class B1 and class C1, as identified in the initial step.}
\label{tab:stats_appendix}
\end{table*}

\section{Clustering in {\it Fermi} + BATSE GRB dataset}
\label{fermi_batse_clustering}
\begin{figure*}
    \centering
\includegraphics[width=\textwidth]{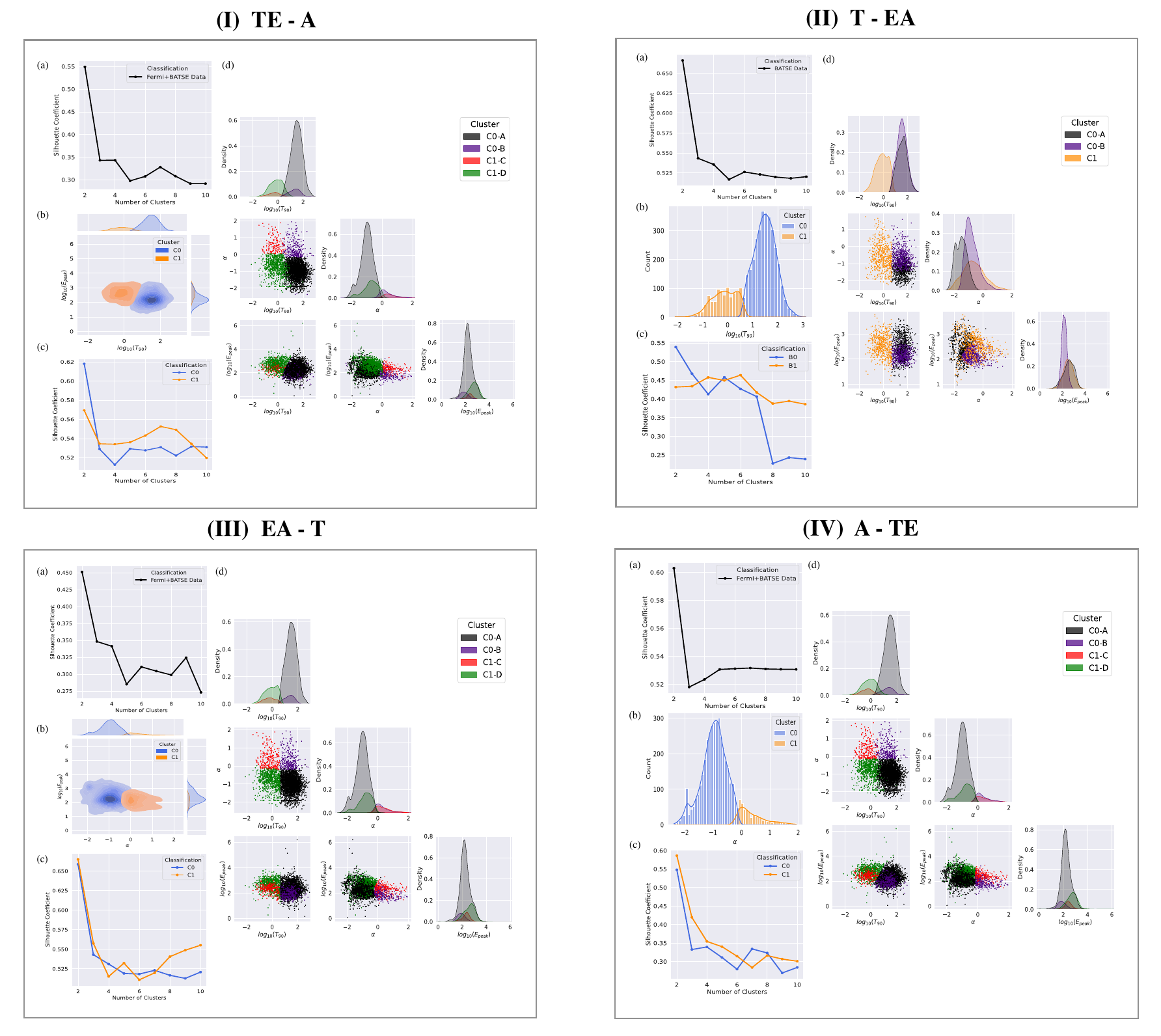}
    \caption{Presented above are the outcomes of the NGMM classification performed on the {\it Fermi} + BATSE GRB dataset. The series of plots labeled I, II, III, and IV showcase the results derived from distinct classification sequences: TE-A, T-EA, EA-T, and A-TE, respectively. Within each plot grouping, (a) and (c) plots illustrate the Silhouette Coefficient (SC) obtained during the initial and subsequent steps of the sequence, correspondingly. While (b) depict the Kernel Density Estimation (KDE) showcasing the distribution of properties for the two classes identified in the first step. In the sequence of T-EA, (d) showcases the sub-classes, C0-A (black) and C0-B (purple) obtained in the second step of classification of C0, while C1 is represented in orange. In all other plot groupings, (d) shows all four classes (C0-A in black, C0-B in purple, C1-C in red and C1-D in green) obtained in the second step of the classification process.}
    \label{Fermi_Batse}
\end{figure*}

The clustering attempt was made on the combined dataset of {\it Fermi} + BATSE. The first step of classification resulted in two classes: C0 (blue) and C1 (orange). All except T-EA sequence resulted in four clusters. In the T-EA sequence, the classification of C1 class in the second step resulted in ambiguity. The results are summarised in Figure \ref{Fermi_Batse} and characteristics of the four classes obtained are listed in the Table \ref{tab:stats_appendix}. 

\newpage
\bibliography{ref}

\begin{thebibliography}{}
\expandafter\ifx\csname natexlab\endcsname\relax\def\natexlab#1{#1}\fi
\providecommand{\url}[1]{\href{#1}{#1}}
\providecommand{\dodoi}[1]{doi:~\href{http://doi.org/#1}{\nolinkurl{#1}}}
\providecommand{\doeprint}[1]{\href{http://ascl.net/#1}{\nolinkurl{http://ascl.net/#1}}}
\providecommand{\doarXiv}[1]{\href{https://arxiv.org/abs/#1}{\nolinkurl{https://arxiv.org/abs/#1}}}

\bibitem[{{Abbott} {et~al.}(2017){Abbott}, {Abbott}, {Abbott}, {Acernese}, {Ackley}, {Adams}, {Adams}, {Addesso}, {Adhikari}, {Adya}, {Affeldt}, {Afrough}, {Agarwal}, {Agathos}, {Agatsuma}, {Aggarwal}, {Aguiar}, {Aiello}, {Ain}, {Ajith}, {Allen}, {Allen}, {Allocca}, {Aloy}, {Altin}, {Amato}, {Ananyeva}, {Anderson}, {Anderson}, {Angelova}, {Antier}, {Appert}, {Arai}, {Araya}, {Areeda}, {Arnaud}, {Arun}, {Ascenzi}, {Ashton}, {Ast}, {Aston}, {Astone}, {Atallah}, {Aufmuth}, {Aulbert}, {et~al.}}]{Abbott2017}
{Abbott}, B.~P., {Abbott}, R., {Abbott}, T.~D., {et~al.} 2017, \apjl, 848, L13, \dodoi{10.3847/2041-8213/aa920c}

\bibitem[{{Ackermann} {et~al.}(2013){Ackermann}, {Ajello}, {Asano}, {Axelsson}, {Baldini}, {Ballet}, {Barbiellini}, {Bastieri}, {Bechtol}, {Bellazzini}, {Bhat}, {Bissaldi}, {Bloom}, {Bonamente}, {Bonnell}, {Bouvier}, {Brandt}, {Bregeon}, {Brigida}, {Bruel}, {Buehler}, {Burgess}, {Buson}, {Byrne}, {Caliandro}, {Cameron}, {Caraveo}, {Cecchi}, {Charles}, {et~al.}}]{Ackermann_etal_2013_LATcatalog}
{Ackermann}, M., {Ajello}, M., {Asano}, K., {et~al.} 2013, \apjs, 209, 11, \dodoi{10.1088/0067-0049/209/1/11}

\bibitem[{{Acuner} \& {Ryde}(2018)}]{Acuner_Ryde2018}
{Acuner}, Z., \& {Ryde}, F. 2018, \mnras, 475, 1708, \dodoi{10.1093/mnras/stx3106}

\bibitem[{{Acuner} {et~al.}(2019){Acuner}, {Ryde}, \& {Yu}}]{Acuner_etal_2020}
{Acuner}, Z., {Ryde}, F., \& {Yu}, H.-F. 2019, \mnras, 487, 5508, \dodoi{10.1093/mnras/stz1356}

\bibitem[{{Ahlgren} {et~al.}(2015){Ahlgren}, {Larsson}, {Nymark}, {Ryde}, \& {Pe'er}}]{Ahlgren_etal_2015}
{Ahlgren}, B., {Larsson}, J., {Nymark}, T., {Ryde}, F., \& {Pe'er}, A. 2015, \mnras, 454, L31, \dodoi{10.1093/mnrasl/slv114}

\bibitem[{Ahumada \& Singer(2021)}]{Ahumada2021}
Ahumada, T., \& Singer, L.~P. 2021, Nature Astronomy, 5, 917, \dodoi{10.1038/s41550-021-01428-7}

\bibitem[{{Ahumada} {et~al.}(2021){Ahumada}, {Singer}, {Anand}, {Coughlin}, {Kasliwal}, {Ryan}, {Andreoni}, {Cenko}, {Fremling}, {Kumar}, {Pang}, {Burns}, {Cunningham}, {Dichiara}, {Dietrich}, {Svinkin}, {Almualla}, {Castro-Tirado}, {De}, {Dunwoody}, {Gatkine}, {Hammerstein}, {Iyyani}, {Mangan}, {Perley}, {Purkayastha}, {Bellm}, {Bhalerao}, {Bolin}, {Bulla}, {Cannella}, {Chandra}, {Duev}, {Frederiks}, {Gal-Yam}, {Graham}, {Ho}, {Hurley}, {Karambelkar}, {Kool}, {Kulkarni}, {Mahabal}, {Masci}, {McBreen}, {Pandey}, {Reusch}, {Ridnaia}, {Rosnet}, {Rusholme}, {Carracedo}, {Smith}, {Soumagnac}, {Stein}, {Troja}, {Tsvetkova}, {Walters}, \& {Valeev}}]{Ahumada_etal_2021}
{Ahumada}, T., {Singer}, L.~P., {Anand}, S., {et~al.} 2021, Nature Astronomy, 5, 917, \dodoi{10.1038/s41550-021-01428-7}

\bibitem[{{Asano} {et~al.}(2009){Asano}, {Inoue}, \& {M{\'e}sz{\'a}ros}}]{Asano_etal_2009}
{Asano}, K., {Inoue}, S., \& {M{\'e}sz{\'a}ros}, P. 2009, \apj, 699, 953, \dodoi{10.1088/0004-637X/699/2/953}

\bibitem[{{Atwood} {et~al.}(2009){Atwood}, {Abdo}, {Ackermann}, {Althouse}, {Anderson}, {Axelsson}, {Baldini}, {Ballet}, {Band}, {Barbiellini}, {Bartelt}, {Bastieri}, {Baughman}, {Bechtol}, {B{\'e}d{\'e}r{\`e}de}, {Bellardi}, {Bellazzini}, {Berenji}, {Bignami}, {Bisello}, {Bissaldi}, {Blandford}, {Bloom}, {Bogart}, {Bonamente}, {Bonnell}, {Borgland}, {Bouvier}, {Bregeon}, {Brez}, {Brigida}, {et~al.}}]{Atwood2009}
{Atwood}, W.~B., {Abdo}, A.~A., {Ackermann}, M., {et~al.} 2009, \apj, 697, 1071, \dodoi{10.1088/0004-637X/697/2/1071}

\bibitem[{{Band} {et~al.}(1993){Band}, {Matteson}, {Ford}, {Schaefer}, {Palmer}, {Teegarden}, {Cline}, {Briggs}, {Paciesas}, {Pendleton}, {Fishman}, {Kouveliotou}, {Meegan}, {Wilson}, \& {Lestrade}}]{Band_etal_1993}
{Band}, D., {Matteson}, J., {Ford}, L., {et~al.} 1993, \apj, 413, 281, \dodoi{10.1086/172995}

\bibitem[{{Belkin} {et~al.}(2024){Belkin}, {Pozanenko}, {Minaev}, {Pankov}, {Volnova}, {Rossi}, {Stratta}, {Benetti}, {Palazzi}, {Moskvitin}, {Burhonov}, {Rumyantsev}, {Klunko}, {Inasaridze}, {Reva}, {Kim}, {Jelinek}, {Kann}, {Volvach}, {Volvach}, {Xu}, {Zhu}, {Fu}, \& {Mkrtchyan}}]{Belkin_etal_2024}
{Belkin}, S., {Pozanenko}, A.~S., {Minaev}, P.~Y., {et~al.} 2024, \mnras, 527, 11507, \dodoi{10.1093/mnras/stad3989}

\bibitem[{{Beloborodov}(2011)}]{Beloborodov2011}
{Beloborodov}, A.~M. 2011, ApJ, 737, 68, \dodoi{10.1088/0004-637X/737/2/68}

\bibitem[{{Beloborodov}(2013)}]{Beloborodov2013}
---. 2013, ApJ, 764, 157, \dodoi{10.1088/0004-637X/764/2/157}

\bibitem[{{Beniamini} {et~al.}(2018){Beniamini}, {Barniol Duran}, \& {Giannios}}]{Beniamini_etal_2018}
{Beniamini}, P., {Barniol Duran}, R., \& {Giannios}, D. 2018, \mnras, 476, 1785, \dodoi{10.1093/mnras/sty340}

\bibitem[{{Burgess} {et~al.}(2020){Burgess}, {B{\'e}gu{\'e}}, {Greiner}, {Giannios}, {Bacelj}, \& {Berlato}}]{Burgess_etal_2020}
{Burgess}, J.~M., {B{\'e}gu{\'e}}, D., {Greiner}, J., {et~al.} 2020, Nature Astronomy, 4, 174, \dodoi{10.1038/s41550-019-0911-z}

\bibitem[{{Burgess} {et~al.}(2015){Burgess}, {Ryde}, \& {Yu}}]{Burgess_etal_2015}
{Burgess}, J.~M., {Ryde}, F., \& {Yu}, H.-F. 2015, \mnras, 451, 1511, \dodoi{10.1093/mnras/stv775}

\bibitem[{Cano {et~al.}(2017)Cano, Wang, Dai, \& Wu}]{Cano2017}
Cano, Z., Wang, S.-Q., Dai, Z.-G., \& Wu, X.-F. 2017, Advances in Astronomy, 2017, 1, \dodoi{10.1155/2017/8929054}

\bibitem[{{Daigne} {et~al.}(2011){Daigne}, {Bo{\v s}njak}, \& {Dubus}}]{Daigne2011}
{Daigne}, F., {Bo{\v s}njak}, {\v Z}., \& {Dubus}, G. 2011, \aap, 526, A110, \dodoi{10.1051/0004-6361/201015457}

\bibitem[{{Della Valle} {et~al.}(2006){Della Valle}, {Chincarini}, {Panagia}, {Tagliaferri}, {Malesani}, {Testa}, {Fugazza}, {Campana}, {Covino}, {Mangano}, {Antonelli}, {D'Avanzo}, {Hurley}, {Mirabel}, {Pellizza}, {Piranomonte}, \& {Stella}}]{GRB060614A_2006}
{Della Valle}, M., {Chincarini}, G., {Panagia}, N., {et~al.} 2006, \nat, 444, 1050, \dodoi{10.1038/nature05374}

\bibitem[{{Dereli-B{\'e}gu{\'e}} {et~al.}(2020){Dereli-B{\'e}gu{\'e}}, {Pe'er}, \& {Ryde}}]{Dereli_etal_2020}
{Dereli-B{\'e}gu{\'e}}, H., {Pe'er}, A., \& {Ryde}, F. 2020, \apj, 897, 145, \dodoi{10.3847/1538-4357/ab9a2d}

\bibitem[{{Dichiara} {et~al.}(2023){Dichiara}, {Tsang}, {Troja}, {Neill}, {Norris}, \& {Yang}}]{Dichiara_etal_2023}
{Dichiara}, S., {Tsang}, D., {Troja}, E., {et~al.} 2023, \apjl, 954, L29, \dodoi{10.3847/2041-8213/acf21d}

\bibitem[{{Dimple} {et~al.}(2023){Dimple}, {Misra}, \& {Arun}}]{Dimple_etal_2023}
{Dimple}, {Misra}, K., \& {Arun}, K.~G. 2023, \apjl, 949, L22, \dodoi{10.3847/2041-8213/acd4c4}

\bibitem[{{Fishman}(2013)}]{Fishman2013}
{Fishman}, G.~J. 2013, in EAS Publications Series, Vol.~61, EAS Publications Series, ed. A.~J. {Castro-Tirado}, J.~{Gorosabel}, \& I.~H. {Park}, 5--14, \dodoi{10.1051/eas/1361001}

\bibitem[{{Fynbo} {et~al.}(2006){Fynbo}, {Watson}, {Th{\"o}ne}, {Sollerman}, {Bloom}, {Davis}, {Hjorth}, {Jakobsson}, {J{\o}rgensen}, {Graham}, {Fruchter}, {Bersier}, {Kewley}, {Cassan}, {Castro Cer{\'o}n}, {Foley}, {Gorosabel}, {Hinse}, {Horne}, {Jensen}, {Klose}, {Kocevski}, {Marquette}, {Perley}, {Ramirez-Ruiz}, {Stritzinger}, {Vreeswijk}, {Wijers}, {Woller}, {Xu}, \& {Zub}}]{Fynbo_etal_2006}
{Fynbo}, J. P.~U., {Watson}, D., {Th{\"o}ne}, C.~C., {et~al.} 2006, \nat, 444, 1047, \dodoi{10.1038/nature05375}

\bibitem[{{Ghisellini} \& {Celotti}(1999)}]{Ghisellini1999}
{Ghisellini}, G., \& {Celotti}, A. 1999, ApJ, 511, L93, \dodoi{10.1086/311845}

\bibitem[{{Giannios}(2008)}]{Giannios2008}
{Giannios}, D. 2008, \aap, 480, 305, \dodoi{10.1051/0004-6361:20079085}

\bibitem[{{Giannios}(2012)}]{Giannios2012}
---. 2012, MNRAS, 422, 3092, \dodoi{10.1111/j.1365-2966.2012.20825.x}

\bibitem[{Goldstein {et~al.}(2013)Goldstein, Preece, Mallozzi, Briggs, Fishman, Kouveliotou, Paciesas, \& Burgess}]{BATSE_catalog_Goldstein_2013}
Goldstein, A., Preece, R.~D., Mallozzi, R.~S., {et~al.} 2013, The Astrophysical Journal Supplement Series, 208, 21, \dodoi{10.1088/0067-0049/208/2/21}

\bibitem[{{Granot} {et~al.}(2000){Granot}, {Piran}, \& {Sari}}]{Granot_etal_2000}
{Granot}, J., {Piran}, T., \& {Sari}, R. 2000, \apjl, 534, L163, \dodoi{10.1086/312661}

\bibitem[{{Gruber} {et~al.}(2014){Gruber}, {Goldstein}, {Weller von Ahlefeld}, {Narayana Bhat}, {Bissaldi}, {Briggs}, {Byrne}, {Cleveland}, {Connaughton}, {Diehl}, {Fishman}, {Fitzpatrick}, {Foley}, {Gibby}, {Giles}, {Greiner}, {Guiriec}, {van der Horst}, {von Kienlin}, {Kouveliotou}, {Layden}, {Lin}, {Meegan}, {McGlynn}, {Paciesas}, {Pelassa}, {Preece}, {Rau}, {Wilson-Hodge}, {Xiong}, {Younes}, \& {Yu}}]{Gruber_GBM_spectral_catalog_2014}
{Gruber}, D., {Goldstein}, A., {Weller von Ahlefeld}, V., {et~al.} 2014, \apjs, 211, 12, \dodoi{10.1088/0067-0049/211/1/12}

\bibitem[{Hakkila {et~al.}(2003)Hakkila, Giblin, Roiger, Haglin, Paciesas, \& Meegan}]{Hakkila2003}
Hakkila, J., Giblin, T.~W., Roiger, R.~J., {et~al.} 2003, The Astrophysical Journal, 582, 320, \dodoi{10.1086/344568}

\bibitem[{{Hakkila} {et~al.}(2000){Hakkila}, {Haglin}, {Pendleton}, {Mallozzi}, {Meegan}, \& {Roiger}}]{Hakkila_etal_2000}
{Hakkila}, J., {Haglin}, D.~J., {Pendleton}, G.~N., {et~al.} 2000, \apj, 538, 165, \dodoi{10.1086/309107}

\bibitem[{Hartigan \& Wong(1979)}]{Hartigan1979}
Hartigan, J.~A., \& Wong, M.~A. 1979, JSTOR: Applied Statistics, 28, 100

\bibitem[{{Horv{\'a}th} {et~al.}(2019){Horv{\'a}th}, {Hakkila}, {Bagoly}, {T{\'o}th}, {R{\'a}cz}, {Pint{\'e}r}, \& {T{\'o}th}}]{Horvath_etal_2019}
{Horv{\'a}th}, I., {Hakkila}, J., {Bagoly}, Z., {et~al.} 2019, \apss, 364, 105, \dodoi{10.1007/s10509-019-3585-1}

\bibitem[{{Iyyani} {et~al.}(2013){Iyyani}, {Ryde}, \& et~al.}]{Iyyani2013}
{Iyyani}, S., {Ryde}, F., \& et~al. 2013, MNRAS, 433, 2739, \dodoi{10.1093/mnras/stt863}

\bibitem[{{Iyyani} \& {Sharma}(2021)}]{Iyyani_Sharma2021}
{Iyyani}, S., \& {Sharma}, V. 2021, \apjs, 255, 25, \dodoi{10.3847/1538-4365/ac082f}

\bibitem[{{Iyyani} {et~al.}(2015){Iyyani}, {Ryde}, {Ahlgren}, {Burgess}, {Larsson}, {Pe'er}, {Lundman}, {Axelsson}, \& {McGlynn}}]{Iyyani_etal_2015}
{Iyyani}, S., {Ryde}, F., {Ahlgren}, B., {et~al.} 2015, \mnras, 450, 1651, \dodoi{10.1093/mnras/stv636}

\bibitem[{Jespersen {et~al.}(2020)Jespersen, Severin, Steinhardt, Vinther, Fynbo, Selsing, \& Watson}]{Jespersen2020}
Jespersen, C.~K., Severin, J.~B., Steinhardt, C.~L., {et~al.} 2020, The Astrophysical Journal, 896, L20, \dodoi{10.3847/2041-8213/ab964d}

\bibitem[{{Kouveliotou} {et~al.}(1993){Kouveliotou}, {Meegan}, {Fishman}, {Bhat}, {Briggs}, {Koshut}, {Paciesas}, \& {Pendleton}}]{Kouveliotou1993}
{Kouveliotou}, C., {Meegan}, C.~A., {Fishman}, G.~J., {et~al.} 1993, \apjl, 413, L101, \dodoi{10.1086/186969}

\bibitem[{{Kovacevic} {et~al.}(2014){Kovacevic}, {Izzo}, {Wang}, {Muccino}, {Della Valle}, {Amati}, {Barbarino}, {Enderli}, {Pisani}, \& {Li}}]{Kovacevic_etal_2014}
{Kovacevic}, M., {Izzo}, L., {Wang}, Y., {et~al.} 2014, \aap, 569, A108, \dodoi{10.1051/0004-6361/201424700}

\bibitem[{{Levan} {et~al.}(2023{\natexlab{a}}){Levan}, {Gompertz}, {Malesani}, {Tanvir}, {Burns}, {Salvaterra}, {Ackley}, {Lamb}, {Fynbo}, {Schneider}, {Jakobsson}, {Izzo}, {Fruchter}, {Watson}, {Kennedy}, {Hjorth}, {Pugliese}, {Bhirombhakdi}, \& {Dhillon}}]{Levan_etal_2023}
{Levan}, A.~J., {Gompertz}, B.~P., {Malesani}, D.~B., {et~al.} 2023{\natexlab{a}}, GRB Coordinates Network, 33569, 1

\bibitem[{{Levan} {et~al.}(2023{\natexlab{b}}){Levan}, {Gompertz}, {Malesani}, {Tanvir}, {Burns}, {Salvaterra}, {Ackley}, {Lamb}, {Fynbo}, {Schneider}, {Jakobsson}, {Izzo}, {Fruchter}, {Watson}, {Kennedy}, {Hjorth}, {Pugliese}, {Bhirombhakdi}, \& {Dhillon}}]{230307A_KN}
---. 2023{\natexlab{b}}, GRB Coordinates Network, 33569, 1

\bibitem[{{Lundman} {et~al.}(2013){Lundman}, {Pe'er}, \& {Ryde}}]{Lundman2013}
{Lundman}, C., {Pe'er}, A., \& {Ryde}, F. 2013, MNRAS, 428, 2430, \dodoi{10.1093/mnras/sts219}

\bibitem[{{MacFadyen} \& {Woosley}(1999)}]{MacFadyen1999}
{MacFadyen}, A.~I., \& {Woosley}, S.~E. 1999, \apj, 524, 262, \dodoi{10.1086/307790}

\bibitem[{{McCully} {et~al.}(2017){McCully}, {Hiramatsu}, {Howell}, {Hosseinzadeh}, {Arcavi}, {Kasen}, {Barnes}, {Shara}, {Williams}, {V{\"a}is{\"a}nen}, {Potter}, {Romero-Colmenero}, {Crawford}, {Buckley}, {Cooke}, {Andreoni}, {Pritchard}, {Mao}, {Gromadzki}, \& {Burke}}]{170817A_KN_red}
{McCully}, C., {Hiramatsu}, D., {Howell}, D.~A., {et~al.} 2017, \apjl, 848, L32, \dodoi{10.3847/2041-8213/aa9111}

\bibitem[{{Meegan} {et~al.}(2009){Meegan}, {Lichti}, {Bhat}, {Bissaldi}, {Briggs}, {Connaughton}, {Diehl}, {Fishman}, {Greiner}, {Hoover}, {van der Horst}, {von Kienlin}, {Kippen}, {Kouveliotou}, {McBreen}, {Paciesas}, {Preece}, {Steinle}, {Wallace}, {Wilson}, \& {Wilson-Hodge}}]{Meegan2009}
{Meegan}, C., {Lichti}, G., {Bhat}, P.~N., {et~al.} 2009, \apj, 702, 791, \dodoi{10.1088/0004-637X/702/1/791}

\bibitem[{{Papathanassiou} \& {Meszaros}(1996)}]{Papathanassiou1996}
{Papathanassiou}, H., \& {Meszaros}, P. 1996, \apjl, 471, L91, \dodoi{10.1086/310343}

\bibitem[{Pedregosa {et~al.}(2011)Pedregosa, Varoquaux, Gramfort, Michel, Thirion, Grisel, Blondel, Prettenhofer, Weiss, Dubourg, Vanderplas, Passos, Cournapeau, Brucher, Perrot, \& Duchesnay}]{scikit-learn}
Pedregosa, F., Varoquaux, G., Gramfort, A., {et~al.} 2011, Journal of Machine Learning Research, 12, 2825

\bibitem[{{Pe'er}(2008)}]{Pe'er2008}
{Pe'er}, A. 2008, ApJ, 682, 463, \dodoi{10.1086/588136}

\bibitem[{{Pe'er} \& {Waxman}(2004)}]{Peer&Waxman2004}
{Pe'er}, A., \& {Waxman}, E. 2004, ApJ, 613, 448, \dodoi{10.1086/422989}

\bibitem[{Perley {et~al.}(2016)Perley, Niino, Tanvir, Vergani, \& Fynbo}]{Perley2016}
Perley, D.~A., Niino, Y., Tanvir, N.~R., Vergani, S.~D., \& Fynbo, J. P.~U. 2016, Space Science Reviews, 202, 111, \dodoi{10.1007/s11214-016-0237-4}

\bibitem[{{Poolakkil} {et~al.}(2021){Poolakkil}, {Preece}, {Fletcher}, {Goldstein}, {Bhat}, {Bissaldi}, {Briggs}, {Burns}, {Cleveland}, {Giles}, {Hui}, {Kocevski}, {Lesage}, {Mailyan}, {Malacaria}, {Paciesas}, {Roberts}, {Veres}, {von Kienlin}, \& {Wilson-Hodge}}]{Fermi_GRB_spec_catalog2021}
{Poolakkil}, S., {Preece}, R., {Fletcher}, C., {et~al.} 2021, \apj, 913, 60, \dodoi{10.3847/1538-4357/abf24d}

\bibitem[{{Preece} {et~al.}(1998){Preece}, {Briggs}, {Mallozzi}, {Pendleton}, {Paciesas}, \& {Band}}]{Preece_etal_1998}
{Preece}, R.~D., {Briggs}, M.~S., {Mallozzi}, R.~S., {et~al.} 1998, \apjl, 506, L23, \dodoi{10.1086/311644}

\bibitem[{Rajaniemi \& Mahonen(2002)}]{Rajaniemi2002}
Rajaniemi, H.~J., \& Mahonen, P. 2002, The Astrophysical Journal, 566, 202, \dodoi{10.1086/337959}

\bibitem[{{Rastinejad} {et~al.}(2022){Rastinejad}, {Gompertz}, {Levan}, {Fong}, {Nicholl}, {Lamb}, {Malesani}, {Nugent}, {Oates}, {Tanvir}, {de Ugarte Postigo}, {Kilpatrick}, {Moore}, {Metzger}, {Ravasio}, {Rossi}, {Schroeder}, {Jencson}, {Sand}, {Smith}, {Ag{\"u}{\'\i} Fern{\'a}ndez}, {Berger}, {Blanchard}, {Chornock}, {Cobb}, {De Pasquale}, {Fynbo}, {Izzo}, {Kann}, {Laskar}, {Marini}, {Paterson}, {Escorial}, {Sears}, \& {Th{\"o}ne}}]{211211A_KN}
{Rastinejad}, J.~C., {Gompertz}, B.~P., {Levan}, A.~J., {et~al.} 2022, \nat, 612, 223, \dodoi{10.1038/s41586-022-05390-w}

\bibitem[{{Rees} \& {Meszaros}(1994)}]{Rees_Meszaros1994}
{Rees}, M.~J., \& {Meszaros}, P. 1994, \apjl, 430, L93, \dodoi{10.1086/187446}

\bibitem[{Reynolds {et~al.}(2009)}]{reynolds2009gaussian}
Reynolds, D.~A., {et~al.} 2009, Encyclopedia of biometrics, 741

\bibitem[{{Rossi} {et~al.}(2020){Rossi}, {Stratta}, {Maiorano}, {Spighi}, {Masetti}, {Palazzi}, {Gardini}, {Melandri}, {Nicastro}, {Pian}, {Branchesi}, {Dadina}, {Testa}, {Brocato}, {Benetti}, {Ciolfi}, {Covino}, {D'Elia}, {Grado}, {Izzo}, {Perego}, {Piranomonte}, {Salvaterra}, {Selsing}, {Tomasella}, {Yang}, {Vergani}, {Amati}, \& {Stephen}}]{Kilonova_search_2020}
{Rossi}, A., {Stratta}, G., {Maiorano}, E., {et~al.} 2020, \mnras, 493, 3379, \dodoi{10.1093/mnras/staa479}

\bibitem[{Rossi {et~al.}(2022)Rossi, Rothberg, Palazzi, Kann, D’Avanzo, Amati, Klose, Perego, Pian, Guidorzi, Pozanenko, Savaglio, Stratta, Agapito, Covino, Cusano, D’Elia, Pasquale, Valle, Kuhn, Izzo, Loffredo, Masetti, Melandri, Minaev, Guelbenzu, Paris, Paiano, Plantet, Rossi, Salvaterra, Schulze, Veillet, \& Volnova}]{Rossi2022}
Rossi, A., Rothberg, B., Palazzi, E., {et~al.} 2022, The Astrophysical Journal, 932, 1, \dodoi{10.3847/1538-4357/ac60a2}

\bibitem[{{Ryde}(2004)}]{Ryde2004}
{Ryde}, F. 2004, ApJ, 614, 827, \dodoi{10.1086/423782}

\bibitem[{{Ryde} \& {Pe'er}(2009)}]{Ryde&Pe'er2009}
{Ryde}, F., \& {Pe'er}, A. 2009, ApJ, 702, 1211, \dodoi{10.1088/0004-637X/702/2/1211}

\bibitem[{{Ryde} {et~al.}(2011){Ryde}, {Pe'er}, {et~al.}}]{Ryde2011}
{Ryde}, F., {Pe'er}, A., {et~al.} 2011, MNRAS, 415, 3693, \dodoi{10.1111/j.1365-2966.2011.18985.x}

\bibitem[{Salmon {et~al.}(2022)Salmon, Hanlon, \& Martin-Carrillo}]{Salmon2022}
Salmon, L., Hanlon, L., \& Martin-Carrillo, A. 2022, Galaxies, 10, 77, \dodoi{10.3390/galaxies10040077}

\bibitem[{{Sari} {et~al.}(1998){Sari}, {Piran}, \& {Narayan}}]{Sari1998}
{Sari}, R., {Piran}, T., \& {Narayan}, R. 1998, ApJL, 497, L17, \dodoi{10.1086/311269}

\bibitem[{Steinhardt {et~al.}(2023)Steinhardt, Mann, Rusakov, \& Jespersen}]{Steinhardt2023}
Steinhardt, C.~L., Mann, W.~J., Rusakov, V., \& Jespersen, C.~K. 2023, The Astrophysical Journal, 945, 67, \dodoi{10.3847/1538-4357/acb999}

\bibitem[{{Tanvir} {et~al.}(2017){Tanvir}, {Levan}, {Gonz{\'a}lez-Fern{\'a}ndez}, {Korobkin}, {Mandel}, {Rosswog}, {Hjorth}, {D'Avanzo}, {Fruchter}, {Fryer}, {Kangas}, {Milvang-Jensen}, {Rosetti}, {Steeghs}, {Wollaeger}, {Cano}, {Copperwheat}, {Covino}, {D'Elia}, {de Ugarte Postigo}, {Evans}, {Even}, {Fairhurst}, {Figuera Jaimes}, {Fontes}, {Fujii}, {Fynbo}, {Gompertz}, {Greiner}, {Hodosan}, {Irwin}, {Jakobsson}, {J{\o}rgensen}, {Kann}, {Lyman}, {Malesani}, {McMahon}, {Melandri}, {O'Brien}, {Osborne}, {Palazzi}, {Perley}, {Pian}, {Piranomonte}, {Rabus}, {Rol}, {Rowlinson}, {Schulze}, {Sutton}, {Th{\"o}ne}, {Ulaczyk}, {Watson}, {Wiersema}, \& {Wijers}}]{170817A_KN_blue}
{Tanvir}, N.~R., {Levan}, A.~J., {Gonz{\'a}lez-Fern{\'a}ndez}, C., {et~al.} 2017, \apjl, 848, L27, \dodoi{10.3847/2041-8213/aa90b6}

\bibitem[{{Tavani}(1996)}]{Tavani1996}
{Tavani}, M. 1996, ApJ, 466, 768, \dodoi{10.1086/177551}

\bibitem[{{Troja} {et~al.}(2019){Troja}, {Castro-Tirado}, {Becerra Gonz{\'a}lez}, {Hu}, {Ryan}, {Cenko}, {Ricci}, {Novara}, {S{\'a}nchez-R{\'a}mirez}, {Acosta-Pulido}, {Ackley}, {Caballero Garc{\'\i}a}, {Eikenberry}, {Guziy}, {Jeong}, {Lien}, {M{\'a}rquez}, {Pandey}, {Park}, {Sakamoto}, {Tello}, {Sokolov}, {Sokolov}, {Tiengo}, {Valeev}, {Zhang}, \& {Veilleux}}]{160821B_kilonova}
{Troja}, E., {Castro-Tirado}, A.~J., {Becerra Gonz{\'a}lez}, J., {et~al.} 2019, \mnras, 489, 2104, \dodoi{10.1093/mnras/stz2255}

\bibitem[{Troja {et~al.}(2022)Troja, Fryer, O’Connor, Ryan, Dichiara, Kumar, Ito, Gupta, Wollaeger, Norris, Kawai, Butler, Aryan, Misra, Hosokawa, Murata, Niwano, Pandey, Kutyrev, van Eerten, Chase, Hu, Caballero-Garcia, \& Castro-Tirado}]{Troja2022}
Troja, E., Fryer, C.~L., O’Connor, B., {et~al.} 2022, Nature, 612, 228, \dodoi{10.1038/s41586-022-05327-3}

\bibitem[{{Uhm} \& {Zhang}(2014)}]{Uhm_Zhang2014}
{Uhm}, Z.~L., \& {Zhang}, B. 2014, Nature Physics, 10, 351, \dodoi{10.1038/nphys2932}

\bibitem[{{Vianello} {et~al.}(2018){Vianello}, {Gill}, {Granot}, {Omodei}, {Cohen-Tanugi}, \& {Longo}}]{Vianello_etal_2018}
{Vianello}, G., {Gill}, R., {Granot}, J., {et~al.} 2018, \apj, 864, 163, \dodoi{10.3847/1538-4357/aad6ea}

\bibitem[{{von Kienlin} {et~al.}(2020){von Kienlin}, {Meegan}, {Paciesas}, {Bhat}, {Bissaldi}, {Briggs}, {Burns}, {Cleveland}, {Gibby}, {Giles}, {Goldstein}, {Hamburg}, {Hui}, {Kocevski}, {Mailyan}, {Malacaria}, {Poolakkil}, {Preece}, {Roberts}, {Veres}, \& {Wilson-Hodge}}]{Fermi_burst_catalog2020}
{von Kienlin}, A., {Meegan}, C.~A., {Paciesas}, W.~S., {et~al.} 2020, \apj, 893, 46, \dodoi{10.3847/1538-4357/ab7a18}

\bibitem[{{Yang} {et~al.}(2022){Yang}, {Ai}, {Zhang}, {Zhang}, {Liu}, {Wang}, {Yang}, {Yin}, {Li}, \& {L{\"u}}}]{Yang_etal_2022}
{Yang}, J., {Ai}, S., {Zhang}, B.-B., {et~al.} 2022, \nat, 612, 232, \dodoi{10.1038/s41586-022-05403-8}

\bibitem[{{Zhang} \& {Yan}(2011)}]{Zhang&Yan2011}
{Zhang}, B., \& {Yan}, H. 2011, ApJ, 726, 90, \dodoi{10.1088/0004-637X/726/2/90}

\end{thebibliography}
\bibliographystyle{aasjournal}



\end{document}